# A Metric Determined by Photon Exchange

Charles Francis

**Abstract:**
The *k*-calculus was advanced by Hermann Bondi as a means of explaining special relativity using only simple algebra [1]. Bondi's argument was placed in the context of classical electrodynamics. In this paper it is placed in context of particle theoretic QED. The central derivations are reviewed, using proofs only slightly more elegant than those in Bondi's books. As used by Bondi, *k* is Doppler red shift. The relativistic measurement of position is placed in the context of an information theoretic interpretation of quantum mechanics as a theory of measurement results. The law of geodesic motion is seen as consequent on the refraction of the wave function due curvature, as in geometric optics. The *k*-calculus is extended to include gravitational red shift and to account for gravity by observing that, after allowing a small inherent delay in the reflection of a photon, the metric defined by the radar method obeys Einstein's field equation. A very simple derivation of Schwarzschild is given using the *k*-calculus and avoiding differential geometry and calculation of Christoffel symbols, and the Newtonian approximation is seen from a direct application of red shift to the wave function.

Charles Francis
Clef Digital Systems Ltd.
Lluest, Neuaddlwyd
Lampeter
Ceredigion
SA48 7RG

15/8/01

## 1  Background

The question as to whether measured distances are a prior property of empty space, or whether they are simply relationships found in matter has been open since the introduction of absolute space in Newton's Principia and the criticisms levelled against it by Leibniz and others. Although the mathematical formulation of physical law has depended on an assumption of space, or more recently space-time, the Leibnizian relationist view (which had previously be advocated by Descartes and has its roots in Aristotle, Democritus and other Greek writers) continues to hold intellectual appeal, and there is some reason both within foundations of quantum mechanics and in relativity for thinking that this would be the correct way to formulate physical theory [3]. In recent years the Leibnizian view has been advocated by Smolin [15], Rovelli [12] and others, as motivation for work on spin networks, spin foams and other background free theories.

Hermann Bondi expressed the view that *'with our modern outlook and modern technology the Michelson-Morley experiment is a mere tautology"* [2], the reason being that reference frames in space-time require light for their definition. This is not quite true, in part because the photon could have (or acquire) a non-zero but immeasurably small mass, but principally because in quantum field theory the amplitude for the creation of a particle and its annihilation at any non-synchronous point is non-zero, even outside the light cone, so the speed of individual photons is not constrained. Nonetheless we can discuss the maximum theoretical speed of information, and we need to do so in order to talk about reference frames or time and space co-ordinates.

If we are to measure the time and distance of an event spacially separated from ourselves, then information must travel between us and the event. If we know the speed of information transfer, we can easily determine the time and distance of the event. But speed is defined in terms of time and distance, which leads to a paradox. The 4-coordinate of an event must be known before we can talk of the speed of information, but the speed of information must be known to determine the 4-coordinate of an event. To resolve the paradox we must find something fundamental, and base everything else on it. If we do not accept instantaneous action at a distance, then we may say that there is always a maximum speed of information, which we can call $c$. Natural units in which $c = 1$ and $\hbar = 1$ will be used in this paper ($G \neq 1$). It is tautologous to say that the maximum speed of information is the same (up to scaling) in all reference frames, because there is no reference frame which does not depend for its definition on the maximum speed of information. In practice light does travel at $c$, to the limits of experimental accuracy, and for the purpose of this paper it will be taken that light is the carrier of information.

In the $k$-calculus the radar method is used to measure of time and distance co-ordinates. This can be taken as the definition of space time co-ordinates since any other method of measurement of the time or position of an event can be calibrated to give an identical result to radar. Radar is preferred to a ruler, because it applies directly to both large and small distances, and because a single measurement can be used for both time and space co-ordinates. Quantum electrodynamics has shown that the exchange of photons is responsible for the electromagnetic force and radar ranging is an instance of photon exchange in the special case that the particle reflecting radar is an eigenstate of position and a space-time diagram showing the reflection of a photon is sharply defined. Since electromagnetic forces are responsible for all the structures of matter in our macroscopic environment, it is not unreasonable to postulate that photon exchange is responsible for geometrical relationships internally within a body as well, by analogy with the process as it takes place in radar.

What is discussed in this paper is an idealisation of radar. It is imagined that a photon can be sent in any given direction at a precise time and that a reflected photon returns after an interval which can be precisely timed. There is no such thing as a perfectly confined wave packet, but, as previously remarked, the definition of time and space coordinates depends on the maximum theoretical speed of information in any direction, not on the practical issues of signalling with e.m radiation. For example the simplest possible antenna, the dipole antenna, has a transmission/receiving pattern that resembles a figure 8. The transmitting pattern introduces an uncertainty in the direction in which photons are transmitted, and there is a corresponding uncertainty in our ability to determine the direction from which a received photon came. The smaller the object we are trying to detect by radar, the more its radiation pattern resembles that of a dipole. Hence, there is a relationship between the uncertainty in the position of an object and the size of the object compared to the wavelength of the probing signal. To reduce uncertainty to achieve perfect eigenstates of position as discussed here would require radar signals of infinitesimal wavelength and infinite energy. It is legitimate to discuss such idealisations since a wave packet is a superposition of eigenstates. Indeed the radar pulse can be thought of as a wave packet describing the uncertainty in time of transmission of a photon. The definition of a metric does not depend on such practical issues but on a bounding value, the maximum theoretical speed of information, which can only be attained in the theoretical limit of a radar pulse of infinitesimal duration. Bondi's $k$-calculus was placed in the context of classical e.m. radiation.

QED is normally formulated with a fixed background Minkowsky metric, but there is no greater problem with the description of wave mechanics or field theory in curved space-time beyond the acknowledge problems of defining field theory in Minkowsky space time [16], and there is a substantial literature on semi-classical solutions in curved space time. Using a mathematical formulation of QED such as Scharf's Finite QED [13] or Discrete QED [9], either of which applies as a good approximation in a locality (i.e. when the locality is near flat and behaviour at infinity is not important), resolves ultraviolet divergence problems by correctly treating Wick's theorem, and specifies that equal time products of field operators are normally ordered so that charge and energy can be written as well defined products of fields, it is reasonable to modify the metric in a simply connected local region containing no singularities. The problem then is to find a physical reason or mechanism to justify the choice of a particular metric. It is the purpose of this paper to present a mechanism based on physical processes found in measurement and thereby to unify graity with electrodynamics, claiming that QED should be formulated as a background free theory with a curved geometry satisfying Einstein's field equation. It is rephrased here in the context of particle theoretic QED in the tradition of Dirac and Feynman (see e.g. [14] for a discussion of views on the interpretation of QED). However a reader may prefer to ignore the interpretational comments in section 3, *Information Space Interpretation*, and treat this paper in a semi-classical manner, as a limit obtained for a classical radar pulse of duration tending to zero. In either interpretation, the evolution of the wave function is treated by geometric optics.

## 2   Special Relativity

There is room for confusion between two very similar questions, 'What is time?' and 'What is the time?'. The first question has something to do with consciousness, and our perception of time as a flow from past to future. It admits no easy answer, but is quite distinct from the second question and only the second question is relevant to physics, or to the definition of space-time co-ordinates. The answer to the question 'What is the time?' is always something like 4:30 or 6:25. *The time is a number read from a clock*. There are many different types of clock, but every clock has two common elements, a repeating process and a counter. The rest of the mechanism converts the number of repetitions to conventional units of time. A good clock should provide accurate measurement and it should give a uniform measure of time. The count is in integer cycles of the process in the clock, so for accurate measurement the process must repeat as rapidly as possible. In a uniform clock, the repeating process must repeat each time identical to the last, uninfluenced by external matter. Clocks may be called identical if they measure identical units of time when they are together, and if moving them does not noticeably affect the physical processes of their operation.

A clock defines the time, but does so only at one place. A space-time co-ordinate system also requires a definition of distance, and a definition of time at a distance from the clock. This is provided for by the radar method. Then the distance of an event is half the lapsed time for light to go from the clock to the event and return to the clock. the time at which the signal is reflected is the mean time between when it is sent and when it returns. The radar method also measures direction; it will be seen from Pythagoras' theorem that the algebra is formally identical for 3-vectors and for one dimensional space-time diagrams. Each point on a space-time diagram represents an event. Space-time diagrams are defined such that lines of equal time are horizontal and lines of equal distance are vertical and light is drawn at $45^o$ (figure 1).

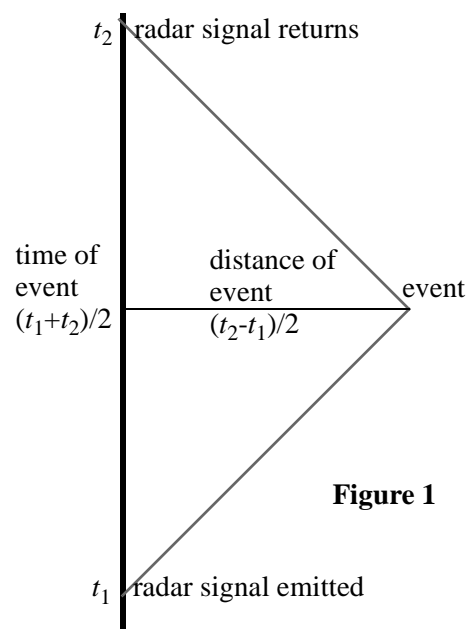

**Figure 1**

By definition uniform motion is shown by a straight line on a space-time diagram. In figure 2, a space craft is uniformly moving in the Earth's reference frame. The space craft and the Earth have identical clocks and communicate with each other by radio or light. The Earth sends the space craft two signals at an interval $t$. The space craft receives them at an interval $kt$ on the space craft's clock. $k \in \mathbb{R}$ is immediately recognisable as red shift (by considering the signals as the start and stop of a burst of light of a set number of wavelengths of a set frequency). Similarly if the observer on the space craft sends two signals at an interval $t$ on his clock, they are received at an interval $k't$ on the Earth. The defining condition for the special theory of relativity is that there is a special class of reference frames such that *red shift is both constant and equal for both observers*, $k = k'$. The general theory of relativity relaxes this condition and results in the force of gravity, but if there is no preferred orientation in space-time then whenever clocks coincide $k = k'$, so that Minkowsky space-time applies locally everywhere. For the remainder of this section it is assumed that $k = k'$, this being the condition for the special, rather than the general, theory of relativity.,

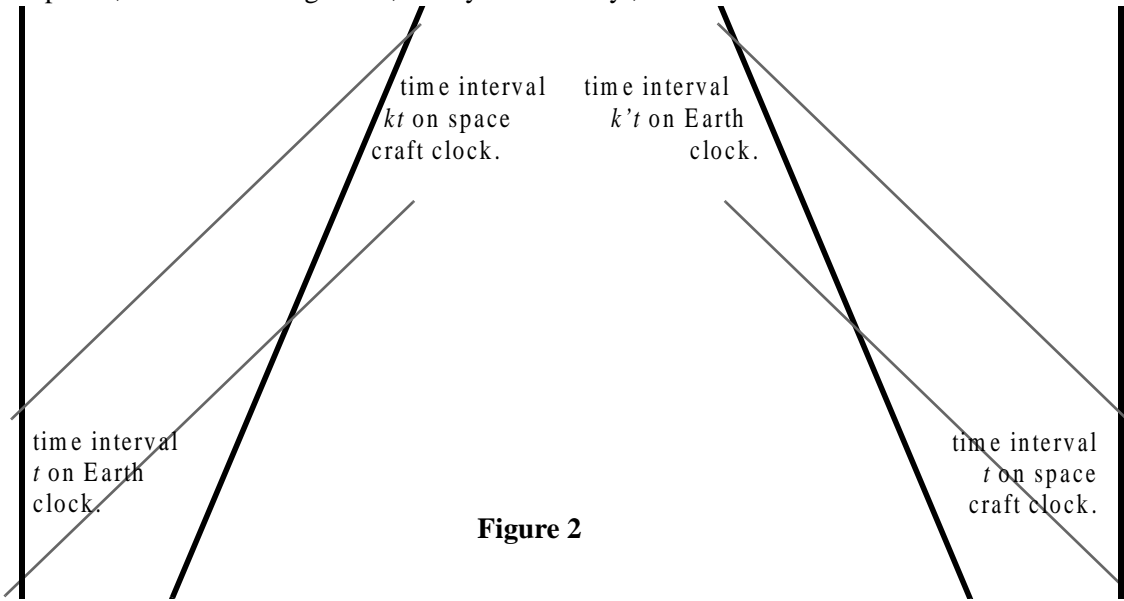

**Figure 2**

**Theorem:** (Time dilation, figure 3) The time $T$ measured by a space craft's clock during an interval $t$ on the Earths clock is given by

2.1 $$T = t\sqrt{1 - v^2}$$

**Proof:** The space craft and the Earth set both clocks to zero at the moment the space craft passes the Earth. The space craft is moving at speed $v$, so by definition, after time $t$ on the Earth clock, the space craft has travelled distance $vt$. Therefore Earth's signal was sent at time $t - vt$, and returned at time $t + vt$. For inertial reference frames, if the space craft sends the Earth signals at an interval $t$ the Earth receives them at an interval $kt$. So

2.2 $$T = k(t - vt).$$

Then by applying the Doppler shift again for the signal coming back

2.3 $$t + vt = k^2(t - vt)$$

Eliminating $k$ gives 2.1, the formula for time dilation.

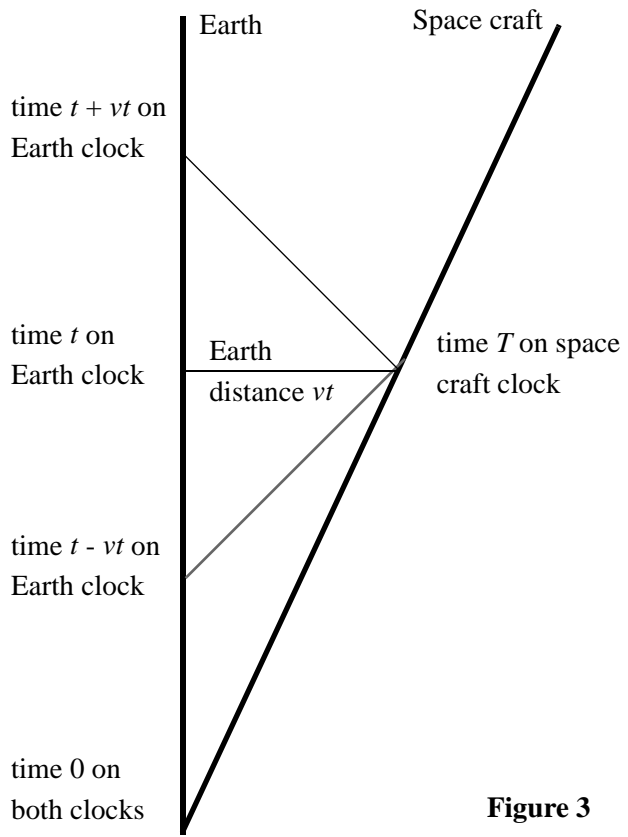

**Figure 3**

**Theorem:** (Lorentz Contraction, figure 4) A distance *d* on the earth is measured on a space craft to be

2.4  $$D = \frac{d}{\sqrt{1-v^2}}$$

**Proof:** The bow and stern of the space craft are shown as parallel lines. The space craft's clock is in the bow. The space craft and Earth set their clocks to zero when the bow passes the Earth clock. Earth uses radar to measure the distance, *d*, to the stern, by sending a signal at time $-d$, which returns at time *d* on the Earth clock. The same signal is used to measure *D* on the spaceship. By the Doppler shift, the outgoing signal passes the bow at time $-(d/k)$ on the space craft's clock, and the returning signal reaches the bow at time *kd*. So, according to the moving space craft

2.5  $$D = (kd + d/k)/2$$

Eliminating *k* using 2.3 gives 2.4.

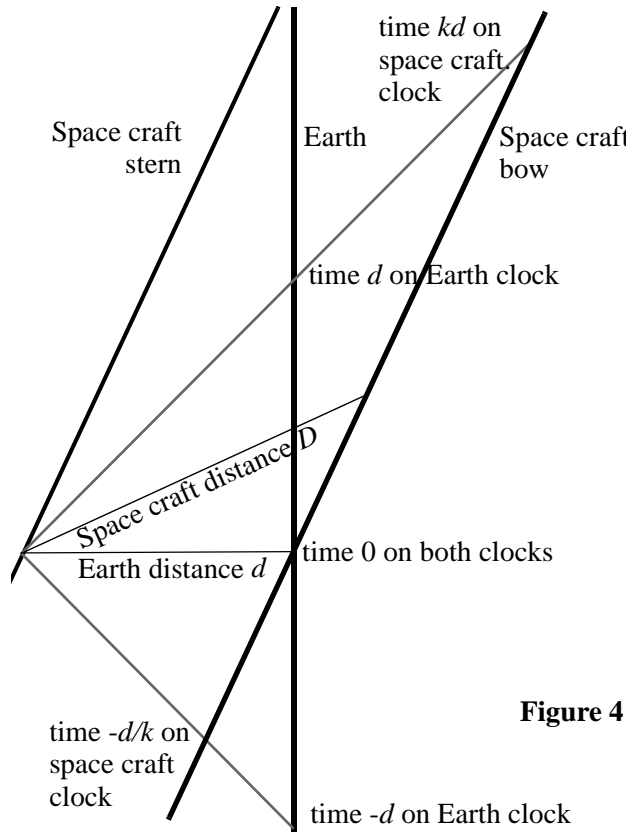

Figure 4

## 3  Information Space Interpretation

We make no assumption of an ontological background in which matter is placed and define a reference frame as the set of potential results of measurement of position. In this view geometry is simply and literally world (*geo-*) measurement (*-metry*); to understand geometry we must study how observers measure space-time co-ordinates. Each observer has a clock, which is, without loss of generality, the origin of his co-ordinate system and which measures proper time for that observer. Each event is given a co-ordinate by measuring the time taken for light to travel to and fro the event. Then the observer's reference frame is defined as the set of possible co-ordinates which could be found in a measurement of position. It is bounded by practical considerations and is part of information space for a particular observer, not prior ontology.

The notion that space-time appears not as an invariant background but as an observer dependent set of potential measurement results is in strict accordance with the orthodox interpretation of quantum mechanics. In Dirac's words "*In the general case we cannot speak of an observable having a value for a particular state, but we can .... speak of the probability of its having a specified value for the state, meaning the probability of this specified value being obtained when one makes a measurement of the observable*" [4]. Taken to its logical conclusion, when this statement is applied to the position observable, it follows that precise position only exists in measurement of position, and hence that there is no ontological background geometrical space or space-time.

In the absence of information the actual configuration of particles cannot be described. In accordance with strict Copenhagen interpretation according to Heisenberg [10] kets are names or labels for states, not descriptions of matter (there are no matter waves and Bohr's notion of complementarity is denied). In a typical measurement in quantum mechanics a particle is studied in near isolation. The suggestion is that there are too few ontological relationships to generate the numerical property of position. Position coordinates do not exist prior to the measurement, and the measurement itself is responsible for introducing interactions which generate position. In this case, prior to the measurement, the state of the system is not labelled by a definite position, but the laws of Hilbert space may be used to define labels containing information about other states – namely the information about what would happen in a measurement. This is done in [8] and [9].

According to this interpretation matter has an unknown but real behaviour which is not directly described by quantum mechanics. The state describes not what is, but what might happen in measurement, and gives the probability for each outcome. The laws of vector space represents a weighted logical OR between possible outcomes of hypothetical measurement in many valued logic. As Rovelli [12] says, *"According to Descartes, there is no 'empty space'. There are only objects and it makes sense to say that an object A is contiguous to object B. The 'location' of an object A is the set of the objects to which A is contiguous. 'Motion' is change in location"*. In the information space interpretation qed describes this motion by the creation/annihilation of particles, and the form of the field operator describes the fact that creation/annihilation might be anywhere; it does not describe a quantised "matter field" which is, in some sense, everywhere. In this formulation the functional integral, or 'sum over all paths' has as natural interpretation, not that a particle passes through all paths in space-time (as described by Feynman for example in [7]), but that the sum over paths is a weighted logical OR between the possible paths that might be detected if an experiment were done to trace the path. Each particle follows a precise but unknown path, where the path is strictly from interaction to interaction with other particles, not through points in space-time. In the case of an eigenstate of definite position measured by the radar method, a refelected photon follows a definite path as shown in figure 1, the probability amplitude for other paths being zero. Interaction takes on Descartes meaning of contiguous, and the structure is embedded in $\mathbb{R}^n$ solely for the purpose of mathematical description. Such an embedding is justified on empirical grounds as a good approximation in a locality.

The measurement of time and position is sufficient for the study of many (it has been said all) other physical quantities; for example a classical measurement of velocity may be reduced to a time trial over a measured distance, and a typical measurement of momentum of a particle involves tracing its track in a bubble chamber. Using the formulation of discrete QED for finiteness [9], Hilbert space is defined using a synchronous space-like co-ordinate system, with time as a parameter as in non-relativistic quantum mechanics. Hilbert space has a basis in measurable values, which are bounded and integral in units of resolution of the apparatus. There is no significance in the bound, $\nu$, of a given co-ordinate system $N \subset \mathbb{N}^3$. If matter goes outside of N it is merely moving out of a co-ordinate system, not out of the universe. Generally it is possible to describe its motion in another co-ordinate system with another origin. Even if it is not intended to take the limit $\nu \to \infty$, N may chosen large enough to say with certainty up to the limit of experimental accuracy, that it contains any particle under study for the duration of the experiment. Matter outside of N is ignored and conservation of conservation of probability is imposed, as usual. N may be curvilinear or rectilinear. A Hilbert space is defined in each reference frame using states $\{|x\rangle | x \in N\}$ as a basis. The manifold will be the collection of possible co-ordinate systems for all possible observers or a continuous superset thereof.

The inner product is frame dependent, as one might expect since it refers to the measurement results of an observer

$$3.1 \qquad \langle g|f\rangle = \sum_{x \in N} \langle g|x\rangle\langle x|f\rangle$$

The definition of a state $|p\rangle$ of definite momentum

$$3.2 \qquad \langle x|p\rangle = e^{ip \cdot x}$$

is replaced by

$$3.3 \qquad \langle x|p\rangle = e^{ipgx}$$

and the discrete Hilbert space is embedded into a vector space in which $p$ and $x$ are continuous variables. The inner product is still defined by 3.1, so that state vectors can be represented by covariant wave functions even though the inner product is observer dependent. The general wave function for a non-interacting particle is

$$3.4 \qquad \forall x \in \mathbb{R}^4 \quad f(x) = \left(\frac{1}{2\pi}\right)^{\frac{3}{2}} \int_M d^3p \, \langle p|f\rangle \, e^{-ipgx}$$

where $M = [-\pi, \pi] \otimes [-\pi, \pi] \otimes [-\pi, \pi] \subset \mathbb{R}^3$ is the 3-torus. The momentum space wave function $\langle p|f\rangle$ is a constant of the motion, as is the vector $p = (E, p)$, where $E = p^0$ is defined by the mass shell condition. Then the discrete probability amplitude for position is found by restricting the domain of 3.4 $x \in N$ and integral bounded $x^0$, for some bound on time such that the probability that the particle leaves N is negligible.

$$3.5 \qquad E^2 = m^2 + p^2$$

The momentum operator is given by the covariant derivative

$$3.6 \qquad P^\nu = -i\nabla^\nu$$

$P^\nu$ is a vector obeying the mass shell condition, 3.5 (or Klein Gordon equation), so that

$$3.7 \qquad P^2 = -g_{\mu\nu}\nabla^\mu \nabla^\nu = m^2$$

is a scalar constant. The bra/ket notation is used for creation and annihilation operators $|x\rangle$ and $\langle x|$ which generate the Fock space of symmetric/antisymmetric states as described in [9] (overloading notation as there is no ambiguity). Then for any real valued functional, $R$, on the space of wave functions which can be written in the form 3.4, we can write a hermitian operator, $R$

$$3.8 \qquad R = \sum_{x \in N} |x\rangle R \langle x|$$

$\langle \underline{x}|$ is the annihilation operator for a particle at $(x) = (x^0, x)$ and $|\overline{x}\rangle$ is the creation operator for the antiparticle (for Dirac particles $\langle \underline{x}| = \langle x|$ at time $x^0$, but this is not true for bosons). Then the field operator is defined by

$$3.9 \qquad \phi(x) = |\overline{x}\rangle + \langle \underline{x}|$$

# A Metric Determined by Photon Exchange    8

Since interactions, including those internal to the structure of a measuring apparatus, are described in terms of local field operators 3.8 can represent only one term of a physical observable operator, which is a normally ordered product of fields

$$3.10 \qquad R = \sum_{x \in N} :\phi^\dagger(x) R \phi(x):$$

In practice we have physical separation of matter and antimatter except during the process or pair creation/annihilation itself, and the rapidly oscillating zitterbewegung terms can be ignored, so that observable operators can be given in the form 3.8, provided also that R can be written as a combination of creation and annihilation operators. For example the (frame dependent) momentum observable is

$$3.11 \qquad P^i = \frac{-i}{2} \sum_{x \in N} |x\rangle [\langle x + 1^i| - \langle x - 1^i|] \text{ for } i = 1, 2, 3$$

So that

$$3.12 \qquad P^i|p\rangle = \frac{-i}{2} \sum_{x \in N} |x\rangle [\langle x + 1^i| - \langle x - 1^i|]|p\rangle = \sum_{x \in N} |x\rangle \langle x|p\rangle \sin p^i = \sin p^i |p\rangle$$

So the eigenvalue of momentum is $\sin p \approx p$ for $p$ much less than the bound of $\pi/\chi$. In the next section we will see is some indication that magnitude of the discrete unit of time for an elementary particle of mass $m$ is $4Gm/c^3$, where $G$ is the gravitation constant. We use $\chi = 4Gm/c^3 = 9.02 \times 10^{-66}$ sec for definiteness. Then an electron with a difference of 0.1% between $p$ and $\sin(p)$ would have an energy of $0.055\pi/\chi = 1.38 \times 10^{52}$ eV, which may be thought unrealistic, and the momentum observable 3.12 is approximated by the vector momentum operator 3.6.

## 4    Gravitational Red Shift

In general distances are defined in terms of time on a given clock and it cannot be assumed that two identical clocks will keep time when separated, even if they are stationary with respect to each other. Moving clocks may be treated by observing that space local to the clock is Minkowsky and applying Lorentz transformation. The net red shift contains both gravitational and Doppler parts. Draw a space-time diagram (figure 5), such that light is drawn at 45° and lines of equal time are horizontal. Just as the observer measures locally Minkowsky co-ordinates $x^\mu$ using his clock as a reference, so locally Minkowsky primed co-ordinates $x^{\mu'}$ can be set up using the distant clock as a reference. The change in speed of the clock is determined by red shift, $k$, (by considering the signals as the start and stop of a burst of light of a set number of wavelengths of a set frequency)

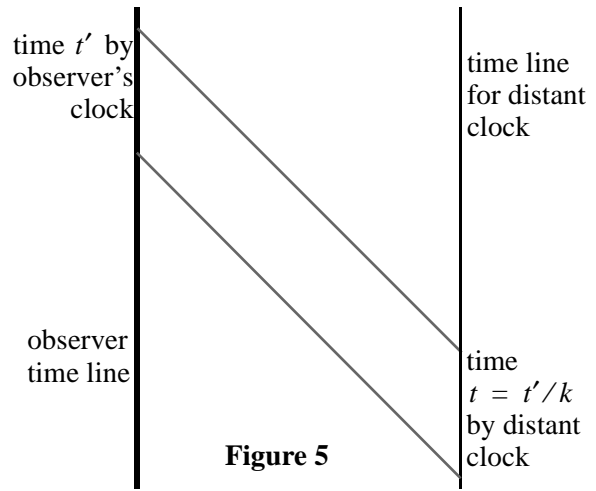

Figure 5

$$4.1 \qquad x^0{}_{,0'} = \frac{\partial x^0}{\partial x^{0'}} = 1/k$$

This is just the time like component of the general transformation law

4.2 $\qquad dx^\mu = x^\mu_{,\nu'} dx^{\nu'}$ with $x^0_{,0'} = 1/k$

So we define the generalisation of red shift

4.3 $\qquad k^\mu_{\nu'} = x^\mu_{,\nu'}$ such that $k^0_{0'} = 1/k$

For any vectors $A^\mu$, $B^\nu$

4.4 $\qquad g_{\alpha'\beta'} A^{\alpha'} B^{\beta'} = g_{\mu\nu} A^\mu B^\nu = g_{\mu\nu} x^\mu_{,\alpha'} x^\nu_{,\beta'} A^{\alpha'} B^{\beta'}$

And since this is true for all values of $A^{\alpha'}$, $B^{\beta'}$ we can infer

4.5 $\qquad g_{\alpha'\beta'} = g_{\mu\nu} x^\mu_{,\alpha'} x^\nu_{,\beta'} = g_{\mu\nu} k^\mu_{\alpha'} k^\nu_{\beta'}$

So the metric in a curved space time can be found from the locally Minkowsky metric of a given observer and generalised red shift 4.3. 4-momentum $p$ obeys the vector transformation law

4.6 $\qquad p^\mu = k^\mu_{\nu'} p^{\nu'}$

which is an invariant for all locally Minkowsky frames, showing that generalised red shift $k^\mu_{\nu'}$ gives the observed changes in frequency and wavelength in the wave function of a particle moving in a gravitational field as expected. It is clear from 4.6 that, starting from any point in space-time, $p$ traces out a geodesic by parallel transport along itself, so that for a macroscopic body in which the wave function is a tightly confined wave packet in both momentum space and co-ordinate space the law of geodesic motion follows immediately.

## 5    Curvature

In the static co-ordinate system shown in figure 5 there may be two causes of gravitational red shift; $k$ may be a function of position or distance in empty space, and $k$ may be directly dependent on the distribution of matter. In either case a tensor equation is required to describe the geometry, and it is convenient to use an equation describing curvature. In the vacuum case $k$ is a function of distance, and from the homogeneity of the vacuum it follows that this must be described by constant curvature, which gives the cosmological constant term in Einstein's Field equation for empty space,

5.1 $\qquad R_{\mu\nu} = \lambda g_{\mu\nu}$

In the present approach the natural motivation for the cosmological constant is that it would make possible a finite closed universe with any amount of missing mass. In this case the cosmological constant would be determined by large scale structure. It will be ignored for the rest of this paper. It is now possible to follow the standard argument that the Einstein tensor is proportional to stress energy

5.2 $\qquad G^{\alpha\beta} = R^{\alpha\beta} - \tfrac{1}{2} g^{\alpha\beta} R = 8\pi G T^{\alpha\beta}$

but the purpose of this paper is to show that Einstein's field equation 5.2 follows for a metric defined by two way light speed, after taking into account a small delay in reflection. Since Einstein's field equation is satisfied by a Schwarzschild geometry it is sufficient to show that the Schwarzschild geometry obtains for a single point particle in an eigenstate of position. Then the operator form of 5.2 holds for one particle, and so it will holds generally by linearity of operators in quantum mechanics.

# A Metric Determined by Photon Exchange 

In contrast to the instantaneous reflection of a classical e.m. wave, the reflection of a photon is treated here as two events, absorption and emission. Ordinarily in QED, emission can occur at any time, before or after the absorption, and it is necessary to integrate over all such possibilities. In the particular case under study there is a measurement and hence an eigenstate of position. In the classical limit the integration gives a superposition of states equivalent to the instantaneous reflection of a classical e.m. wave. In terms of the path integral, other paths contribute nothing to the amplitude and on the remaining path the photon is instantaneously reflected. According to the interpretation given above, in an eigenstate of position, the photon follows a definite, known path, which may be described by space-time diagrams. If the reflection of a photon could be really instantaneous then photon exchange would give a fixed metric, which may be Minkowsky, or it may be a metric with constant curvature given by 5.1. Clearly this is not true for our universe, so it seems reasonable to look at ways of modifying the analysis.

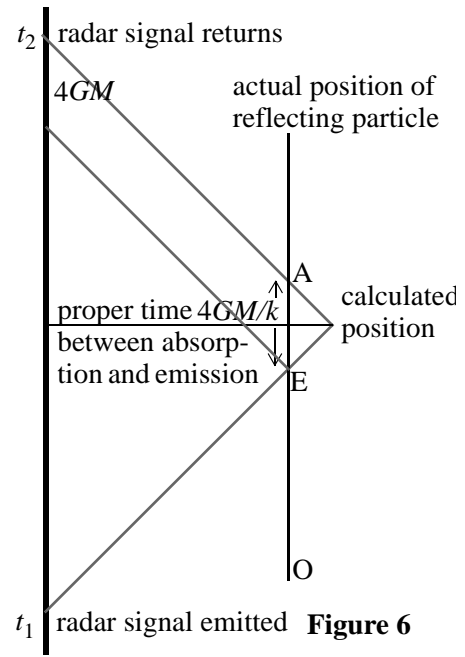

**Figure 6**

A natural modification is to hypothesise a characteristic delay between absorption and emission in proper time of a particle (typically an electron) reflecting a photon (figure 6) . Indeed it has often been suggested that such a small scale correction could justify the cut-off in the treatment of loop divergences and resolve the problem of the Landau pole, and if particle interactions are discrete it is reasonable to anticipate a small delay. The proposed lag is a characteristic of the interaction between elementary particles, and applies to macroscopic phenomena like radar only as as the resultant behaviour of many elementary particle interactions. The metric is determined by the maximum theoretical speed of information, so it is natural to assume that the appropriate lag is the minimum possible time between absorption and emission, rather than the actual time for reflection of a given photon, or even as the expectation of the time of reflection. In this case it is the fundamental unit of discrete time described in [9].

A single elementary particle in an eigenstate of position has spherical symmetry and space-time diagrams may be used to show radial distance without loss of generality. In figure 6 the reflection consists of absorption, A, and emission, E. A and E share the same space-time coordinate, so that the reflection appears instantaneous in a frame defined by radar. This would indicate that the reference frame defined by radar is degenerate in a manner analogous to that of the event horizon of a black hole, where an exterior observer also considers time has stopped. The distance scale for such a degeneracy is the Schwarzschild radius $2GM$ of the electron, or other reflecting particle, corresponding to proper time $4GM/k$ between the interactions. This distance scale gives the scalar curvature in due to elementary particle interactions.

For a perfect eigenstate of position this would be a naked singularity, but in practice there is always uncertainty in position, so that the eigenstate $|x\rangle$ of exact position is replaced with the state $|f\rangle$ for which the probability amplitude for finding the particle at a particular position is $\langle x|f\rangle$. For a scalar particle the corresponding energy density operator is $P^0 = -i\nabla^0$ and the energy density is $\rho = \langle f|x\rangle P^0 \langle x|f\rangle$. Since this is a near eigenstate with the particle at rest we have $P^i = 0$ for $i = 1, 2, 3$ and scalar curvature

5.3 $$R \propto \langle f|x\rangle G P^0 \langle x|f\rangle = G\rho$$

So Einstein's Field Equation is satisfied in the rest frame. The general form of the field equation is found by composing a tensor equation which reduces to 5.3 in the rest frame of a particle. Since all physical properties are comprised of fundamental interactions and hence of field operators, the stress energy tensor operator $T^{\alpha\beta}$ must be a product of field operators, and for a Dirac particle $T^{\alpha\beta}$ has the form

5.4 $$T^{\alpha\beta} = -i{:}\hat{\psi}(x)\gamma^\alpha \nabla^\beta \psi(x){:}$$

where normal ordering is used to avoid an undefined product of fields. It will be observed that experiments to detect gravitational mass also determine the position of a gravitating body, resulting in a reduction of the wave packet; there is no conflict in information space between curvature dependent on energy-density of the wave function, and that dependent on actual mass distribution, as there might be if the manifold was describable as an ontological entity, rather than simply as a set of potential and actual relationships.

When a photon transmits energy from one position to another, curvature is transmitted with it. The natural conclusion is that photonic energy also generates curvature. The appropriate tensor operator is

5.5 $$T^{\alpha\beta} = -i{:}\nabla^\beta A^\alpha(x){:}$$

The full stress energy tensor is the direct sum of operators 5.4, 5.5 for each elementary particle. Then Einstein's field equation takes the form

5.6 $$G^{\alpha\beta} = 8\pi G \langle T^{\alpha\beta}\rangle$$

## 6 Schwarzschild

We seek only to analyse the gravitational effect of one elementary particle, as shown as O in figure 6. Choose radial coordinates with the gravitating particle at the origin, so that $k^\mu_{\nu'}$ is diagonal. The unprimed co-ordinates derive from the proper time of the gravitating particle, whereas the primed co-ordinates pertain to the observer. Draw a synchronous slice through the gravitating particle (figure 7), and superimpose the primed and an unprimed co-ordinate systems. The observer translates his coordinates so that the origins coincide. In the primed reference frame of the observer at radius $r'$ the particle appears as a sphere of radius $2GM$, but in the diagram it is

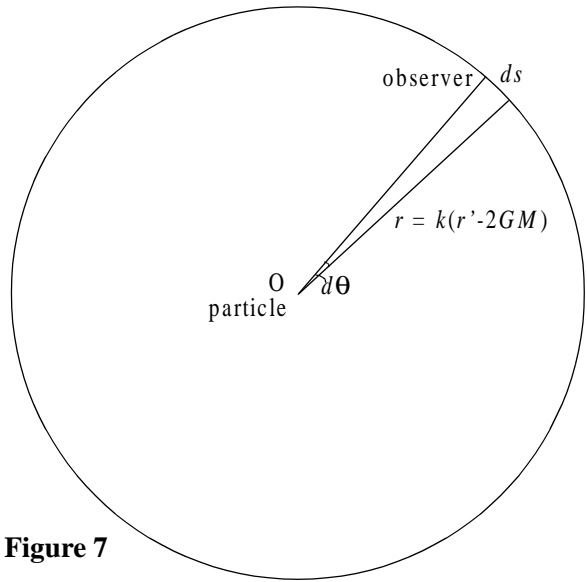

Figure 7

# A Metric Determined by Photon Exchange

squeezed to a point, so we stretch the distance to the observer to compensate. The respective distances from particle to observer, as determined by proper time for the particle and observer, are related by the red shift factor $k$ (as defined in figure 5), so $k$ is the stretch applied to this distance in the primed (observer) co-ordinates.

6.1 $\qquad r = k(r' - 2GM)$

Hence

6.2 $\qquad k = \dfrac{dr}{dr'} = x^1{}_{,1'} = k^1_{1'}$

So that using 4.3 and noting that due to spherical symmetry there is no red shift associated with rotation

6.3 $\qquad k^\mu_\nu = \begin{bmatrix} 1/k & 0 & 0 & 0 \\ 0 & -k & 0 & 0 \\ 0 & 0 & 1 & 0 \\ 0 & 0 & 0 & 1 \end{bmatrix}$

Consider an infinitesimal length $ds$ along the circumference, measured by a clock at the circumference. The clock at the circumference measures red shift $k$ for light coming from the origin, so it is speeded up by that factor, and distances measured by proper time along the circumference are correspondingly shorter. Hence in the unprimed co-ordinates

6.4 $\qquad ds = rd\theta / k$

but in the observer frame, using 6.1,

6.5 $\qquad ds = r'd\theta = k(r - 2GM)d\theta$

By 6.4 and 6.5

6.6 $\qquad k = \left(1 - \dfrac{2GM}{r}\right)^{-1/2}$

Using 4.5 and 6.6 we find the familiar form of the Schwarzschild metric,

6.7 $\qquad g_{\alpha'\beta'} = g_{\mu\nu} k^\mu_{\alpha'} k^\nu_{\beta'} = \begin{bmatrix} \left(1 - \dfrac{2GM}{r}\right) & 0 & 0 & 0 \\ 0 & -\left(1 - \dfrac{2GM}{r}\right)^{-1} & 0 & 0 \\ 0 & 0 & -r^2 & 0 \\ 0 & 0 & 0 & -r^2 \sin^2\theta \end{bmatrix}$

Red shift affects the frequency of the wave function, and hence the energy of a particle, according to the relation

6.8 $\qquad k^0_{0'} p^{0'} = \text{const}$

So the classical energy $E$ satisfies

6.9 $\qquad E = \langle P^0 \rangle \propto 1/k^0_{0'} = k = \left(1 - \dfrac{2GM}{r}\right)^{-1/2} \approx 1 + \dfrac{GM}{r}$

For a body of mass *m* the constant of proportionality is fixed at $r = \infty$

6.10 $$E = m + \frac{GMm}{r}$$

showing the gravitational potential in the Newtonian approximation

**Acknowledgements**

I should like to thank a number of physicists who have discussed the content and ideas of this paper on usenet, particularly Eric Forgy, Mike Mowbray, Frank Wappler, Chris Hillman and John Baez and the moderators of sci.physics.research (John Baez, Matt McIrvin, Ted Bunn & Philip Helbig) for their vigilence in pointing out lack of clarity in expression.